\def\dint{\mathop{\displaystyle \int}}

\newcommand\strt[1]{\rule[-#1pt]{0pt}{#1pt}}

\newcommand\eq{=&&\hspace{-18pt}}

\documentclass[12pt,a4paper]{article}
\usepackage{amsmath}
\usepackage{amssymb}
\usepackage{graphicx}
\usepackage{mathpazo}
\usepackage[usenames,dvipsnames]{xcolor}
\usepackage{hyperref}
\usepackage{multimedia}
\usepackage{comment}
\usepackage[small,nohug,heads=vee]{diagrams}
\diagramstyle[labelstyle=\scriptstyle]
\definecolor{light-gray}{gray}{0.80}

\usepackage{fancyhdr}
\renewcommand{\maketitle}{
    \begin{center}
      \Large
        {\bf Pair production in classical Stueckelberg-Horwitz-Piron electrodynamics}
        \vskip .3 true cm
      \small
        Martin Land \\
        \vskip .3 true cm
        Department of Computer Science \\
        Hadassah College \\
        37 HaNevi'im Street, Jerusalem \\
email: martin@hadassah.ac.il
      \end{center}
      \vskip .5 true cm
}
\input{tcilatex}

\begin{document}
\title{}
\author{}
\maketitle


%
\begin{abstract}

In this paper we calculate pair production from bremsstrahlung as a classical effect in
Stueckelberg-Horwitz-Piron electrodynamics.  In this framework, worldlines are traced
out dynamically through the evolution of events $x^\mu(\tau)$ parameterized by a
chronological time $\tau$ that is independent of the spacetime coordinates.  These
events, defined in an unconstrained 8D phase space, interact through five
$\tau$-dependent gauge fields induced by the event evolution.  The resulting
theory differs in its underlying mechanics from conventional electromagnetism,
but coincides with Maxwell theory in an equilibrium limit.  In particular, the
total mass-energy-momentum of particles and fields is conserved, but the
mass-shell constraint is lifted from individual interacting events, so that the
standard
Feynman-Stueckelberg interpretation of pair creation/annihilation is implemented
in classical mechanics. 


We consider a three-stage interaction which when parameterized by the laboratory
clock $x^0$ 
appears as (1) particle-1 scatters on a heavy nucleus to produce
bremsstrahlung, (2) the radiation field produces a particle/antiparticle pair,
(3) the antiparticle is annihilated 
with particle-2 in the presence of a second heavy nucleus.  When parameterized
in chronological time $\tau$, the underlying process develops as (1) particle-2
scatters on
the second nucleus and begins evolving backward in time with negative
energy, (2) particle-1 scatters on the first nucleus and releases 
bremsstrahlung, (3) particle-2 absorbs radiation which
returns it to forward time evolution with positive energy.

\end{abstract}

\parindent=0cm \parskip=10pt

\section{Introduction}

In the historical introduction to his book on quantum field theory
\cite{weinberg}, Weinberg devotes a paragraph to deprecation of Dirac's hole
theory of antiparticles, observing that QFT had made the theory "unnecessary,
even though it lingers on in many textbooks."  It might also have been mentioned
that the essential idea of hole theory also lingers on productively as the
quasiparticle formalism in condensed matter physics and many-particle
theory.\footnote{More generally, Dirac's fundamental insight that the absence of
a physical object can behave like the presence of an inverse object has been
influential in many fields, especially psychology, and can be compared to the remark attributed to
Miles Davis that, ``Music is the space between the notes. It's not the notes you
play; it's the notes you don't play.''} Still, as an interpretation of
particle/antiparticle processes, the Feynman-Stueckelberg time reversal
formalism, expressed in QFT through the Feynman propagator, has many conceptual
advantages.  One advantage not mentioned by Weinberg is that besides not
requiring the Dirac sea, it does not actually require quantum field theory.  The
description of an antiparticle as a particle propagating backward in time was
first proposed by Stueckelberg \cite{Stueckelberg} in the context of classical
relativity, without resort to quantum ideas or phenomena.  In this model, a pair
process is represented by a single worldline, generated dynamically by a
classical event whose time coordinate advances or retreats with respect to the
laboratory clock, as its instantaneous energy changes sign under interaction
with a field.    

In order to generate worldlines of this type, Stueckelberg constructed a covariant
Hamiltonian theory of interacting spacetime events, in which the events evolve
dynamically, as functions of a Poincar\'{e} invariant parameter $\tau$.  As shown in
Figure 1, the particle worldline is traced out in terms of the values taken on by the
four-vector $x^{\mu }\left( \tau \right) $ as the parameter proceeds monotonically
from $\tau =-\infty $ to $\tau =\infty $.  

\begin{center}
\includegraphics[width=3.5in]{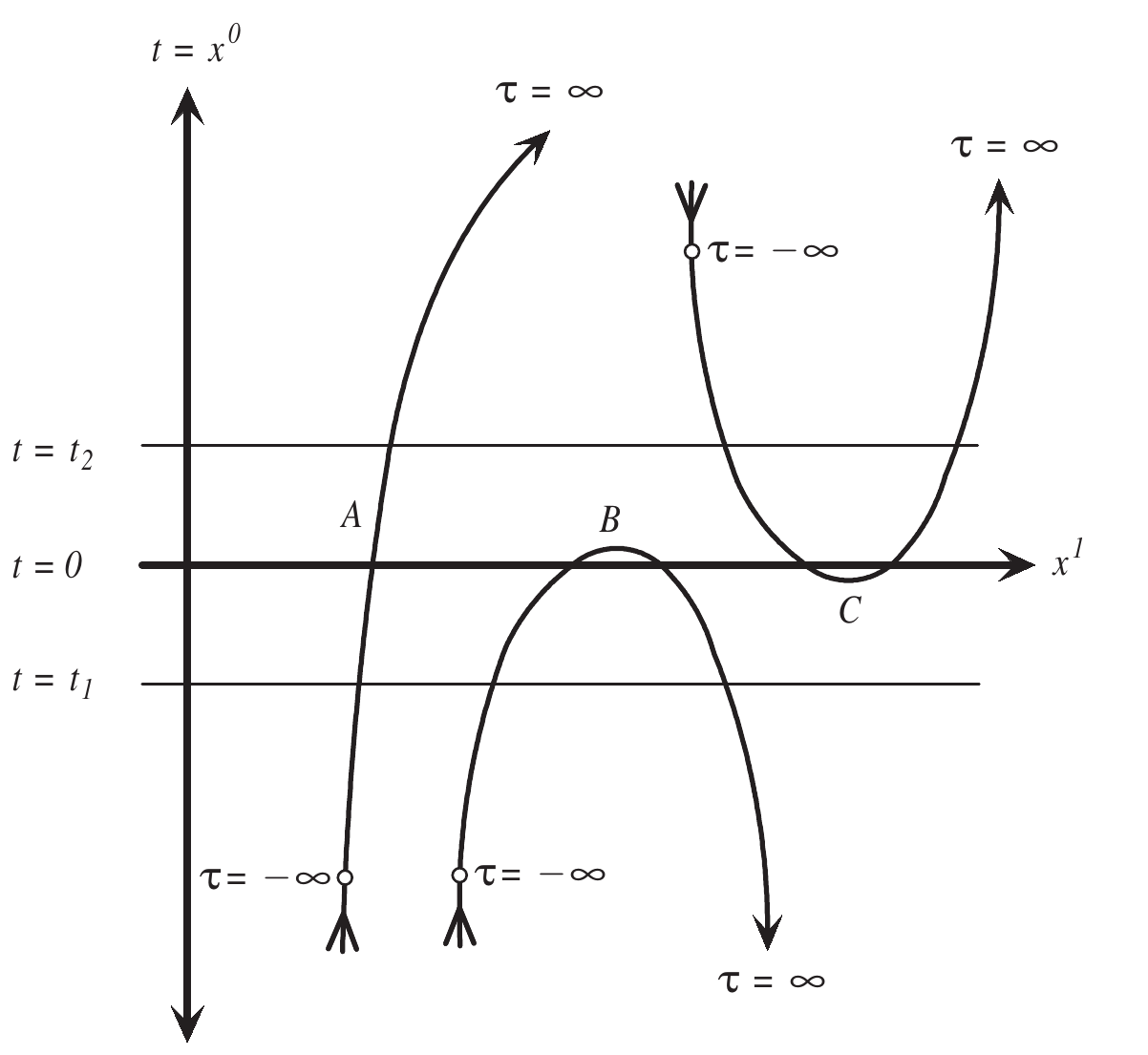}\\
\begin{tabular}{l}
{\footnotesize \ \textbf{Figure 1}: World Lines \cite{Stueckelberg}} \\ 
\begin{tabular}{l}
{\footnotesize A: Usual type, with a unique solution to }$t\left( \tau
\right) =x^{0}${\footnotesize \ for each }$x^{0}$ \\ 
{\footnotesize B: Annihilation type, with two solutions to }$t\left( \tau
\right) =x^{0}${\footnotesize \ for }$x^{0}\ll 0${\footnotesize \ and no
solution for }$x^{0}\gg 0$ \\ 
{\footnotesize C: Creation type, with two solutions to }$t\left( \tau
\right) =x^{0}${\footnotesize \ for }$x^{0}\gg 0${\footnotesize \ and no
solution for }$x^{0}\ll 0$ 
\end{tabular}%
\end{tabular}
\end{center}

By explicitly distinguishing the Einstein coordinate time $x^{0}=t$ from the
temporal order \cite{Two-Aspects}, the parameter time $\tau $ becomes formally
similar to the Galilean invariant time in Newtonian theory, serving
Stueckelberg's broader goal of generalizing the techniques of nonrelativistic
classical and quantum mechanics to covariant form.  Stueckelberg identified pair
creation in worldlines of type C in Figure 1, because there are two solutions to
$t(\tau )=t_{2}$, but no solution to $t(\tau )=t_{1}$. The observer will
therefore first encounter no particle trajectories and then encounter two. It
seems clear that
the intrinsic electric charge should not change along the worldline, but to
identify one part of the worldline as an antiparticle trajectory, 
requires that the measured charge reverse sign. Charge reflection may be grasped
intuitively as in Dirac's hole model: carrying positive charge in one time
direction should be equivalent to carrying negative charge in the opposite time
direction.  In standard QFT, charge conjugation is accomplished through the
action of an operator with no classical analog.  In Stueckelberg's classical
formalism, the $0$-component of the current includes the electric charge multiplied
by $dx^{0}/d\tau $, which becomes negative when the event $x^{\mu }\left( \tau
\right) $ evolves toward earlier values of $t=x^{0}$.  Thus, particles and
antiparticles do not appear as distinct classes of solutions to a defining
equation, but as a single event whose qualitative behavior depends
instantaneously on the dynamical value of its velocity. 

A standard technique for pair creation in the laboratory is the two-step process by
which Anderson \cite{anderson} first observed positrons in 1932: high energy
electrons are first scattered by heavy nuclei to produce bremsstrahlung radiation,
and electron/positron pairs are then created from the radiation field.  The
Bethe-Heitler mechanism \cite{BH} describes this technique as the quantum process,
\begin{eqnarray}
e^- + Z \longrightarrow e^- + &&\mbox{\hspace{-20pt}}Z + \gamma \notag \\
&&\mbox{\hspace{-20pt}}Z + \gamma \longrightarrow Z + e^- + e^+ \notag
\end{eqnarray}
involving a quantized radiation field and the external Coulomb field of the nuclei.  
The Feynman diagrams describing the second step in QED are shown in Figure 2
\cite{lotstedt}.  
\begin{center}
\includegraphics[width=5in]{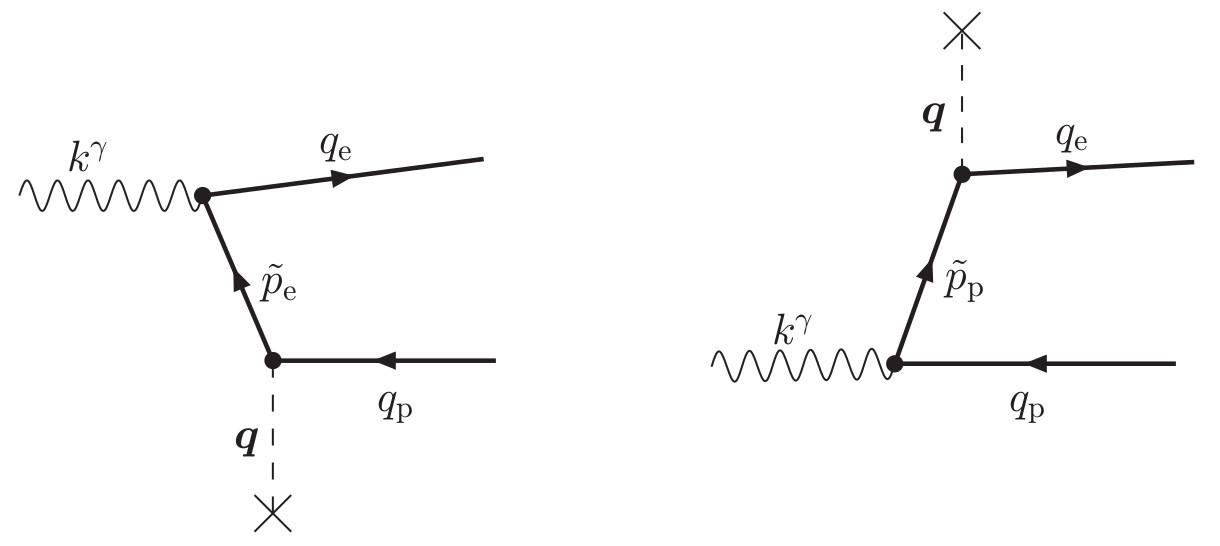}\\
\begin{tabular}{l}
{\footnotesize \ \textbf{Figure 2}: Bethe-Heitler mechanism in QED} $\hspace{7cm}$\\ 
\end{tabular}%
\end{center}
The incoming real photon carries
4-momentum $k^\gamma$ and the real outgoing electron and positron carry
4-momentum $q_e$ and $q_p$.  The intermediate virtual state carries 4-momentum
$\tilde{p}_e$ or $\tilde{p}_p$, and the virtual photon exchanged with the
nucleus is represented by the dashed line.  

A modern experimental setup is shown in Figure 3 \cite{muller}. 
\begin{center}
\includegraphics[width=3.7in]{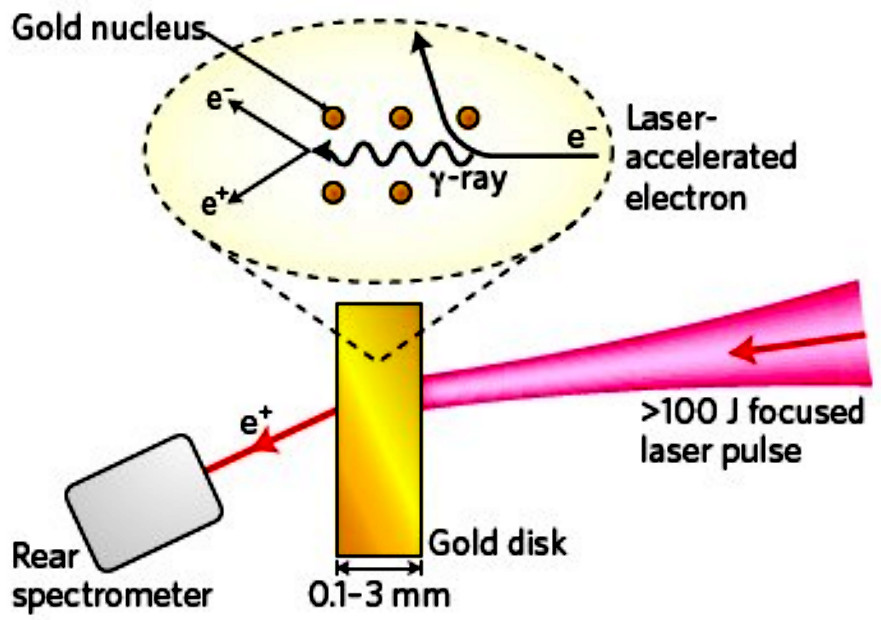}\\
\begin{tabular}{l}
{\footnotesize \ \textbf{Figure 3}: Bethe-Heitler mechanism in the laboratory \cite{muller}} $\hspace{5cm}$\\ 
\end{tabular}%
\end{center}
Electrons are accelerated to high energy by focusing an intense laser pulse on a thin
gold disk.  The electrons are strongly deflected in the Coulomb field of the nuclei
and emit bremsstrahlung represented in the illustration as $\gamma$-rays.  In the
Coulomb field of a nucleus, the photon can decay into an electron/positron pair. 

Stueckelberg was not able to provide a classical account of pair processes, because
the mass-shell constraint $p^2 = (M\dot x)^2 = -M^2$ prevents continuous
evolution of the event trajectory from the timelike region into the spacelike region
on its way to time-reversed timelike motion.  He considered adding a vector component
to his Lorentz force that would overcome the constraint, but dropped the idea, finding
no justification from first principles.  Just such a vector field appears naturally
in a gauge-invariant approach to Stueckelberg's theory \cite{H-refs,saad}.  

In this paper, we give a brief overview of Stueckelberg-Horwitz electrodynamics and use
the formalism to provide a classical description of the pair creation process
described by the Bethe-Heitler mechanism.  Our goal is to calculate the classical
trajectories that produce the two-step process shown in Figure 4.  
\begin{center}
\includegraphics[height=3.5in]{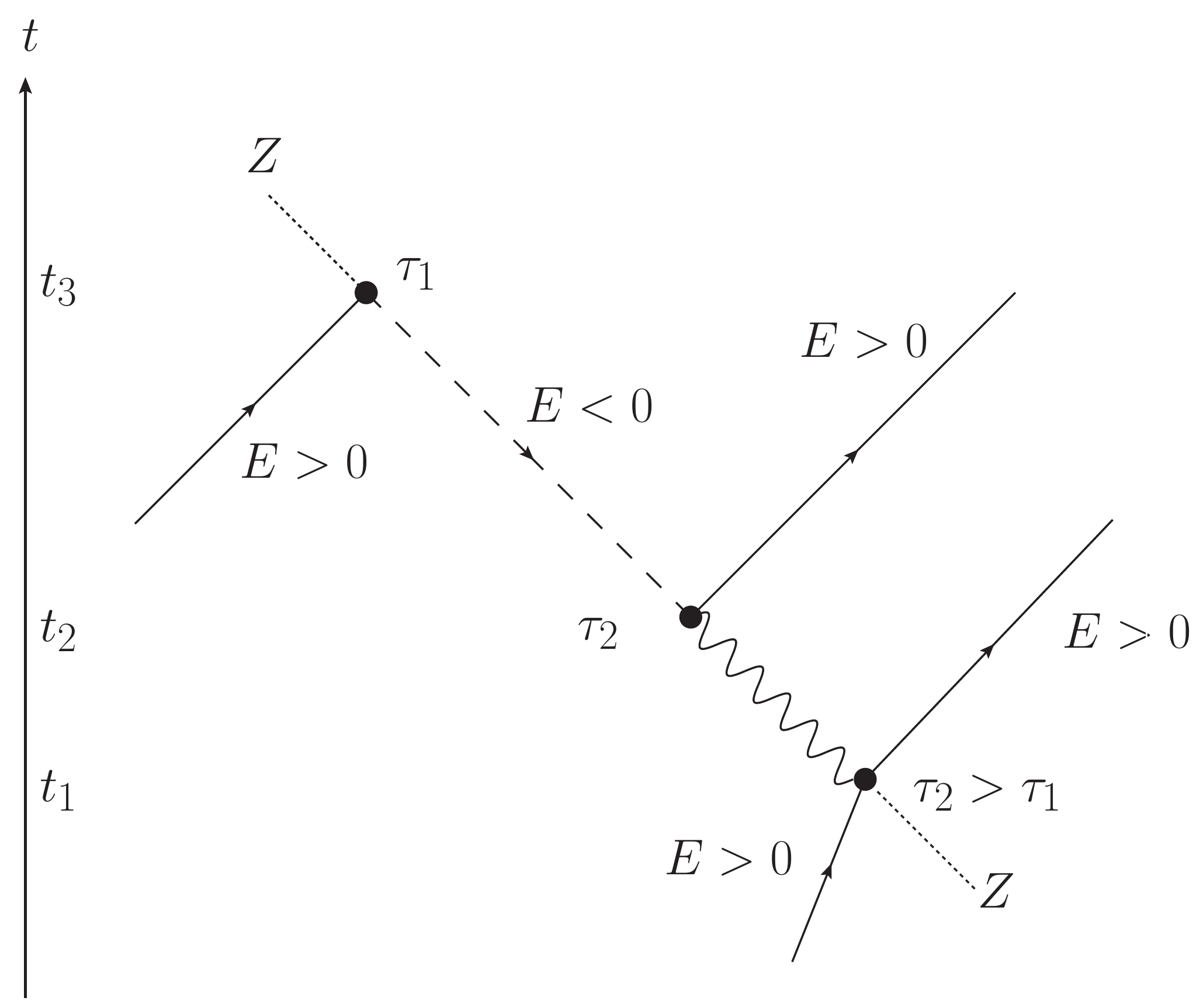}\\
\begin{tabular}{l}
{\footnotesize \ \textbf{Figure 4}: Bethe-Heitler mechanism in classical
electrodynamics} $\hspace{4cm}$\\ 
\end{tabular}%
\end{center}
At chronological time $\tau_1$ positive energy particle-1 arrives at the
laboratory time coordinate $t=t_3$ and scatters in the Coulomb field of nucleus
$Z$.  Particle-1 emerges with negative energy moving backward in $t$.  At a
subsequent time $\tau_2 > \tau_1$ positive energy particle-2 arrives at $t=t_1$
and scatters to positive energy in the field of another nucleus, emitting
classical bremsstrahlung.  This radiation impinges on particle-1 at $t=t_2$, providing
sufficient energy to divert it back to positive energy evolution.  

In the laboratory, where events are recorded in the order determined by clock $t$, the
process appears as particle-2 scattering at $t=t_1$ and emitting bremsstrahlung,
followed by the appearance at $t=t_2$ of a particle/antiparticle pair.  Then at
$t = t_3$, the antiparticle encounters another particle causing their mutual
annihilation. 

In section 2 we present those features of Stueckelberg-Horwitz electrodynamics
required to describe this scattering process.  In section 3 we provide a description
of Coulomb scattering required for the events at $t=t_3$ and $t=t_1$.  In section 4
we calculate the acceleration of a particle in the radiation field of a scattering
event, providing a description of the pair creation event.  In both the pair
creation event at $t=t_2$ and the pair annihilation event at $t=t_3$, the basic
requirement is that the classical interaction energy be greater than the masses of
the created particles.  Section 5 presents a discussion of the results and directions for
further study. 

\section{Overview of Horwitz-Stueckelberg electrodynamics}

The generalized Stueckelberg-Schrodinger equation
\begin{equation}
(i\partial _{\tau }+e_0 \phi)\psi (x,\tau )
=\dfrac{1}{2M}(p^{\mu}-e_0a^{\mu })(p_{\mu }-e_0a_{\mu })\psi (x,\tau )
\label{schr}
\end{equation}
describes the interaction of an event characterized by the wavefunction
$\psi(x,\tau)$ with five gauge fields $a_\mu(x,\tau)$ and $\phi(x,\tau)$.  Equation
(\ref{schr}) is invariant under local gauge transformations
\begin{equation}
\begin{array}{lrcl}
&\psi (x,\tau )&\rightarrow& e^{\, ie_{0}\Lambda (x,\tau )} \, \psi (x,\tau
)\vspace{8pt} \\
\text{Vector potential }\quad&a_{\mu }(x,\tau )&\rightarrow& a_{\mu }(x,\tau
)+\partial _{\mu }\Lambda (x,\tau )\vspace{8pt} \\
\text{Scalar potential }\quad&\phi(x,\tau )&\rightarrow& \phi(x,\tau )+\partial _{\tau }\Lambda (x,\tau ) \\
\end{array}
\end{equation}
whose $\tau$-dependence is the essential departure from Stueckelberg's work, and
determines the structure of the resulting theory \cite{H-refs,saad}.  The corresponding
global gauge invariance leads to the conserved Noether current  
\begin{equation}
\partial _{\mu }j^{\mu }+\partial _{\tau }\rho = 0
\label{cons}
\end{equation}
where
\begin{equation}
j^{\mu }=-\tfrac{i}{2M}\Bigl\{\psi ^*(\partial ^{\mu }-ie_0a^{\mu })\psi -\psi (\partial ^{\mu }+ie_0a^{\mu
})\psi ^*\Bigr\}\qquad \rho =\Bigl|\psi (x,\tau )\Bigr|^{2} \ \ .
\end{equation}
Adopting the formal designations
\begin{equation}
x^5 = \tau \quad\qquad   \quad\qquad  \partial_5 = \partial_\tau \quad\qquad  
\quad j^5 = \rho \quad  \quad\qquad  a_5 =  \phi
\end{equation}
and the index convention 
\begin{equation}
\lambda ,\mu ,\nu =0,1,2,3 \qquad \qquad \alpha ,\beta ,\gamma =0,1,2,3,5
\end{equation}
the gauge and current conditions can be written
\begin{equation}
a_{\alpha }\rightarrow a_{\alpha }+\partial _{\alpha }\Lambda 
\qquad \qquad 
\partial _{\alpha }j^{\alpha } = 0 \ \ .
\end{equation}
The classical mechanics of a relativistic event is found by rewriting the 
Stueckelberg-Schrodinger equation in the form
\begin{equation}
i\partial _{\tau }\psi (x,\tau ) = \left[\dfrac{1}{2M}(p-e_0a)^2 -
e_0a_{5}\right]\psi (x,\tau ) = K \psi (x,\tau ) 
\end{equation}
and transforming the classical Hamiltonian to Lagrangian as
\begin{equation}
L=\dot{x}^{\mu }p_{\mu }-K=\dfrac{1}{2}M\dot{x}^{\mu }\dot{x}_{\mu }+e_0
\dot{x}^{\alpha }a_{\alpha }
\end{equation}
from which the Euler-Lagrange equations
\begin{equation}
\dfrac{d}{d\tau }\dfrac{\partial L}{\partial \dot{x}_{\mu }}
-\dfrac{\partial L}{\partial x_{\mu }}=0
\end{equation}
are
\begin{equation}
\dfrac{d}{d\tau } \Big[ M\dot{x}^{\mu } + e_0 a^{\mu } (x,\tau )\Big]
= e_0 \dot{x}^{\alpha }\partial^\mu a_{\alpha }(x,\tau )
\end{equation}
leading to the Lorentz force 
\begin{equation}
M\ddot{x}^{\mu } = e_0 \big[ \dot{x}^{\alpha} \partial^\mu a_\alpha - (\dot{x}^\nu \partial_\nu + \partial_\tau) a^{\mu } \big]
= e_0 f^\mu_{\; \;  \alpha} (x,\tau )\dot{x}^\alpha
\label{lorentz}
\end{equation}
where
\begin{equation}
f^\mu_{\; \;  \alpha} = \partial^\mu a_\alpha - \partial_\alpha a^\mu
\mbox{\qquad\qquad} \dot{x}^5 = \dot{\tau} = 1 \ \ .
\end{equation}
As required for time reversal, particles may exchange mass with fields
\begin{equation}
\frac{d}{d\tau }(-\tfrac{1}{2}M\dot{x}^{2})=-M\dot{x}^{\mu }\ddot{x}_{\mu
}=-e_0\; \dot{x}^{\mu }(f_{\mu 5}+f_{\mu \nu }\dot{x}^{\nu })=-e_0\; \dot{x
}^{\mu }f_{\mu 5}
\end{equation}
and in this formalism, the mass shell is demoted from the status of constraint to
that of conservation law for interactions in which $f_{\mu 5}=0$.  Analysis of
the mass-energy-momentum tensor shows that the total mass, energy, and momentum
of the particles and fields are conserved.  To write an
electromagnetic action requires the choice of a kinetic term for the gauge field,
which must be both gauge and O(3,1) invariant.  We write
\begin{equation}
S_\text{em} = \int d^{4}xd\tau \left\{e_0j^{\alpha }(x,\tau )a_{\alpha }(x,\tau )-
\int ds\, \frac{\lambda}{4} \left[f^{\alpha \beta }(x,\tau )\Phi (\tau -s)f_{\alpha \beta
}\left( x,s\right) \right] \right\}
\label{action}
\end{equation}
where the local event current 
\begin{equation}
j^\alpha (x,\tau) =  \dot{X}^\alpha(\tau) \delta^4\left(x-X(\tau) \strt{3} \right)
\label{pm_curr}
\end{equation}
has support at the instantaneous location $X(\tau)$ of the event.  The
$\tau$-integral of (\ref{pm_curr}) along the worldline concatenates the event current
into the Maxwell particle current in the usual form.  Taking the field interaction kernel 
to be \cite{H-refs}
\begin{equation}
\Phi (\tau ) = \delta \left( \tau \right) -\lambda ^{2}\delta ^{\prime \prime
}\left( \tau \right) = \int \frac{d\kappa}{2\pi} \,\left[ 1+\left( \lambda
\kappa \right) ^{2}\right] \,e^{-i\kappa \tau } \ \ ,
\label{kernel}
\end{equation}
the inverse function becomes
\begin{equation}
\varphi (\tau ) = \Phi^{-1} = \int \frac{d\kappa}{2\pi}
\,\frac{e^{-i\kappa \tau }}{
1+\left( \lambda \kappa \right) ^{2}}=\frac{1}{2\lambda }e^{-|\tau |/\lambda }
\label{inv}
\end{equation}
which will be seen to spread the current along the worldline.
The classical action can be written using (\ref{kernel}) as
\begin{equation}
S = \int d\tau \; \frac{1}{2}M\dot{x}^{\mu }\dot{x}_{\mu }
+ \int d^4x d\tau \left\{ e_0 \, a_\alpha  j^\alpha 
-\frac{\lambda }{4}f_{\alpha \beta }f^{\alpha \beta }
-\frac{\lambda ^{3}}{4}\left(\partial _{\tau }f^{\alpha \beta
}\right)\left(\partial _{\tau }f_{\alpha \beta }\right)
\right\}
\end{equation}
in which the gauge and O(3,1) invariance are manifest.  The $\tau$ derivatives
in the last term explicitly break any formal higher symmetry in the electromagnetic terms.  
Varying the action in the form (\ref{action}) with respect to the fields 
and applying (\ref{inv}) leads to the field equations
\begin{eqnarray}
&&\partial _{\beta }f^{\alpha \beta }\left( x,\tau \right)
= \frac{e_{0}}{\lambda }\int ds~\varphi \left( \tau -s\right)
j^{\alpha }\left( x,s\right)
= e\, j_\varphi^{\alpha } \left( x,\tau \right) \strt{22}\\
\label{gauss}
&&\partial _{\alpha }f_{\beta \gamma } + 
\partial _{\gamma }f_{\alpha \beta } + 
\partial _{\beta }f_{\gamma \alpha }  = 0 
\end{eqnarray}
which are formally similar to 5D Maxwell equations with $e = e_0 / \lambda$.  The source
$j_\varphi^{\alpha } \left( x,\tau \right)
$
of the field in (\ref{gauss}) is the instantaneous current $j^{\alpha } \left( x,\tau
\right) $ defined in (\ref{pm_curr}) with its support along the worldline smoothed by
the kernel function $\varphi(\tau)$.  For $\lambda$ very small, $\varphi$ becomes a
delta function which narrows the source to a small neighborhood around the event
inducing the current.  The parameter $\lambda$ plays the role of a correlation
length, characterizing the range of the electromagnetic interaction.  

Rewriting the field equations in vector and scalar components, they take the form
\vspace{4pt}
\begin{equation}
\begin{array}{lcl}
\partial _{\nu }\;f^{\mu \nu }- \partial _{\tau }\;f^{5\mu }=e \; j^{\mu }_\varphi 
& \mbox{\qquad} &\partial _{\mu }\;f^{5\mu }=e\;  j^5_\varphi
= \; e  \rho_\varphi
\vspace{8pt}\\ 
\partial _{\mu }f_{\nu \rho }+\partial _{\nu }f_{\rho \mu }+\partial _{\rho }f_{\mu \nu }=0 
& &\partial _{\nu }f_{5\mu }-\partial _{\mu }f_{5\nu }
+ \partial _{\tau }f_{\mu \nu }=0 \\
\end{array}
\label{premax}
\vspace{4pt}
\end{equation}
which may be compared with the 3-vector form of Maxwell equations
\begin{equation}
\vspace{4pt}
\begin{array}{lcl}
\nabla \times \mathbf{B}-\partial _{0}\mathbf{E}=e\mathbf{J} 
& \mbox{\qquad} \mbox{\qquad} & \nabla \cdot \mathbf{E}=eJ^{0} \vspace{8pt}\\ 
\nabla \cdot \mathbf{B}=0
& &\nabla \times \mathbf{E}+\partial _{0}\mathbf{B}=0 \\
\end{array}
%
\end{equation}
with $f_{5\mu }$ playing the role of the vector electric field and $f^{\mu \nu }$
playing the role of the magnetic field.  It is sometimes notationally convenient to
further expand the field into 3-vector components as
\begin{equation}
(\mathbf{e})^i = f^{0i} \qquad (\mathbf{h})_{in} = \epsilon_{ijk} f^{jk} \qquad
(\mathbf{f^5})^i = f^{5i} \ \ .
\end{equation}
The connection with Maxwell theory is found, as seen for the instantaneous event
current, by concatenation --- integration over $\tau$ along the worldline, 
\vspace{4pt}
\begin{equation}
\left. 
\begin{array}{c}
\partial _{\beta }f^{\alpha \beta }\left( x,\tau \right)
=ej_{\varphi}^{\alpha }\left( x,\tau \right) \\ 
\\ 
\partial _{\lbrack \alpha }f_{\beta \gamma ]}=0 \\
\\ 
\partial _{\alpha }j^{\alpha } = 0
\end{array}
\right\}
\underset{\int d\tau }{\mbox{\quad}\xrightarrow{\hspace*{1cm}}
\mbox{\quad}}\left\{ 
\begin{array}{c}
\partial _{\nu }F^{\mu \nu }\left( x\right) =eJ^{\mu }\left( x\right) \\ 
\\ 
\partial _{\lbrack \mu }F_{\nu \rho ]}=0 \\
\\ 
\partial _{\mu } J^{\mu}(x) = 0
\end{array}
\right.
\end{equation}
where
\begin{equation}
%
A^{\mu }(x)=\int d\tau \;a^{\mu }(x,\tau )
\qquad F^{\mu \nu }(x)=\int d\tau \;f^{\mu \nu }(x,\tau )
\qquad J^{\mu}(x) = \int d\tau \; j^{\mu }(x,\tau ) \ \ .
\end{equation}
The field equations (\ref{premax}) are called pre-Maxwell equations, and together
with the Lorentz force (\ref{lorentz}) describe 
a microscopic event dynamics for which Maxwell theory can be understood as an equilibrium limit. 
Since $e_0a^\mu$ must have the dimensions of $eA^\mu$, it follows that 
$e_{0}$ and $\lambda$ have the dimension of time and $e=e_{0}/\lambda$ is dimensionless.
The pre-Maxwell equations lead to the wave equation 
\begin{equation}
\partial _{\alpha }\partial ^{\alpha }a^{\beta }\left( x,\tau \right)
=\left( \partial _{\mu }\partial ^{\mu }- \partial _{\tau }^{2}\right)
a^{\beta }\left( x,\tau \right) =-ej_{\varphi}^{\beta }\left( x,\tau \right)
\end{equation}
whose solutions may respect 5D symmetries broken by the O(3,1) symmetry of the event
dynamics.  The principal part Green's function is
\begin{equation}
\begin{array}{rl}
G(x,\tau )&\hspace{-08pt}=-{\dfrac{1}{{2\pi }}}\delta (x^{2})\delta (\tau )-{\dfrac{1}{{2\pi
^{2}}}\dfrac{\partial }{{\partial {x^{2}}}}}\ {\dfrac{{\theta ( x^{2}-  \tau ^{2})
}}{\sqrt{{ x^{2}-  \tau ^{2}}}}} \; , \mbox{\qquad} x^2 = x^{\mu}x_{\mu}\vspace{08pt}\\
&\hspace{-08pt}=-D\left( x\right) \delta (\tau)-G_{correlation}(x,\tau )
\\
\end{array}
\end{equation}
where $D(x)$ is the 4D Maxwell Green's function and $G_{correlation}$ vanishes under
concatenation.  In this paper we neglect the correlation term.
The `static' Coulomb potential in this framework is induced by an
isolated event moving uniformly along the $t$ axis.  Writing the event as
\begin{equation}
x\left( \tau \right) =\left(  \tau ,0,0,0\right)
\end{equation}
produces the currents
\begin{eqnarray}
j^0(x,\tau) \eq j^5 (x,\tau) =  \delta(t-\tau) \, \delta^4(x) \qquad \qquad {\mathbf j}(x,\tau) = 0 \\
j_\varphi^0(x,\tau) \eq  j_\varphi^5 (x,\tau) =  \varphi(t-\tau) \, \delta^4(x) \qquad \qquad {\mathbf j}_\varphi(x,\tau) = 0
\end{eqnarray}
inducing the Yukawa-type potential
\begin{equation}
a^{0} (x,\tau) =  a^{5}(x,\tau) 
={\frac{e}{{4\pi \vert \mathbf{x} \vert }}}\varphi
\left(\tau - \left( t - \vert \mathbf{x} \vert \right) \right)  
\label{yukawa}
\end{equation}
and recovering the standard Coulomb potential 
\begin{equation}
A^{0}(x)=\dint d\tau \, a^{0}\left( x,\tau \right) = {\dfrac{e}{{4\pi \vert \mathbf{x} \vert}}}
\end{equation}
under concatenation.  A test event on the lightcone of this event will experience the
force
\begin{equation}
M\mathbf{\ddot{x}}
=e^{2}\;\nabla \left[ 
\dfrac{e^{-\left\vert \mathbf{x}\right\vert /\lambda }}{{4\pi }\left\vert 
\mathbf{x}\right\vert }\right]
\end{equation}
in which $1 / \lambda$ represents the mass spectrum of the pre-Maxwell field.  If
$\lambda$ is small (so that $\varphi$ approaches a delta function and the current
narrows to around the event) the mass spectrum becomes wide.  If $\lambda$ is large,
the support of the current spreads along the worldline and the potential becomes
Coulomb-like. 

An arbitrary event $X^{\mu }\left( \tau \right) $ induces the current
\begin{equation}
j_{\varphi }^{\mu }\left( x,\tau \right) = \int ds~\varphi \left( \tau-s\right) 
\dot{X}^{\alpha }\left( s\right) \delta ^{4}\left[ x-X\left(
s\right) \right]
\end{equation}
leading to the Li\'{e}nard-Wiechert potential
\begin{eqnarray}
a^{\beta }\left( x,\tau \right) &&\mbox{\hspace{-20pt}}=-e\int
d^{4}x^{\prime }d\tau ^{\prime }D\left( x-x^{\prime }\right) \delta \left(
\tau -\tau ^{\prime }\right) j_{\varphi }^{\mu }\left( x^{\prime },\tau^{\prime } \right) \\
&&\mbox{\hspace{-20pt}}=\frac{e}{2\pi }\int ds~\varphi \left( \tau
-s\right) \dot{r}^{\alpha }\left( s\right) \delta \left( \left( x-X\left(
s\right) \right) ^{2}\right) \theta ^{ret}
\end{eqnarray}
Using identity
\begin{equation}
\dint d\tau f\left( \tau \right) \delta \left[g\left( \tau \right) \right]
=\dfrac{f\left( \tau_R\right) }{\left\vert g^{\prime }\left( \tau_R\right)
\right\vert } \ \ ,
\end{equation}
where $\tau_R$ is the retarded time found from
\begin{equation}
g\left( \tau \right) =(x-X(\tau_R))^2 =0 \qquad \qquad
\theta ^{ret}=\theta \left( x^{0}-X^{0}\left( \tau_R\right) \right) \ \ ,
\end{equation}
provides
\begin{equation}
a^{\beta }\left( x,\tau \right) =\frac{e}{4\pi }\varphi \left( \tau -\tau_R\right) \frac{
\dot{X}^{\beta }\left( \tau_R\right) }{\left( x^{\mu }-X^{\mu }\left( \tau_R\right)
\right) \dot{X}_{\mu }\left( \tau_R\right)} \ \ .
\end{equation}
Notice that the $\tau$-dependence is limited to the smoothing kernel $\varphi \left( \tau -\tau_R\right)$
and again $\lambda$ plays the role of a correlation length that localizes the
interaction to the neighborhood $\tau_R \pm \lambda $.  Using this potential and writing
\begin{equation}
u^\mu = \dot{X}^\mu (\tau) \qquad z^\mu = x^\mu - X^\mu(\tau)
\end{equation}
we find the field strengths, separated into the retarded and radiation parts, as
\begin{eqnarray}
f_{ret}^{\mu \nu }(x,\tau) \eq -e\varphi \left( \tau -\tau_R\right) \frac{\left( z^{\mu
}u^{\nu }-z^{\nu }u^{\mu }\right) u^{2}}{{4\pi }\left( u\cdot z\right) ^{3}}
\sim \dfrac{1}{\mathbf{z}^{2}} \strt{18}
\\
f_{ret}^{5\mu }(x,\tau)  \eq e\varphi \left( \tau
-\tau_R\right) \frac{z^{\mu }u^{2}-u^{\mu }\left( u\cdot z\right) }{{4\pi }\left(
u\cdot z\right) ^{3}}
\sim \dfrac{1}{\mathbf{z}^{2}}\strt{30}
\\
f_{rad}^{\mu \nu }(x,\tau) \eq -e\varphi (\tau -\tau_R)\frac{\left( z^{\mu }\dot{u}^{\nu
}-z^{\nu }\dot{u}^{\mu }\right) \left( u\cdot z\right) -\left( z^{\mu
}u^{\nu }-z^{\nu }u^{\mu }\right) \left( \dot{u}\cdot z\right) }{{4\pi }%
\left( u\cdot z\right) ^{3}} \notag
\\
&&-{\frac{e}{{4\pi }}}\left[ \frac{z^{\mu }u^{\nu
}-z^{\nu }u^{\mu }}{\left( u\cdot z\right) ^{2}}\right] \frac{d}{d\tau_R}\varphi
(\tau -\tau_R)\sim \dfrac{1}{\left\vert 
\mathbf{z}\right\vert }\strt{24} \label{rad-1}
\\
f_{rad}^{5\mu }(x,\tau)  \eq e\varphi \left( \tau
-\tau_R\right) \frac{\left( \dot{u}\cdot z\right) z^{\mu }}{{4\pi }\left( u\cdot
z\right) ^{3}}-e\frac{z^{\mu }-u^{\mu }\left( u\cdot z\right) }{{%
4\pi }\left( u\cdot z\right) ^{2}}\frac{d}{d\tau_R}\varphi \left( \tau -\tau_R\right)
\sim \dfrac{1}{\left\vert \mathbf{z}\right\vert }
\label{rad-2}
\end{eqnarray}

\section{Coulomb scattering}

We begin by analyzing the scattering at $\tau_1$ in Figure 4.  Initially (at time
$\tau \rightarrow -\infty$) the target nucleus $Z$ and incoming particle are widely
separated.  The nucleus is at rest in the laboratory frame,  
\begin{equation}
X_{Z}\left( \tau \right) =\left( t_Z,{\mathbf x}_Z\right) =\left( 1,{\mathbf 0}\right)
\tau
\end{equation}
and the incoming particle approaches on the trajectory
\begin{equation}
X_{in}\left( \tau \right) 
= \left(t,x,y,z\right) = u\tau +s = \dot{t}_{in} \left( 1,v,0,0\right) \tau
+\left( s_{t},0,s_{y},0\right) 
\end{equation}
where
\begin{equation}
u = \frac{d}{d\tau} \left( t,x,y,z\right)
\qquad\qquad  \frac{dx}{d\tau} =  \frac{dx}{dt} \ \dot{t}_{in} = v \dot{t}_{in}  \qquad \qquad 
\dot{t}_{in} = \frac{dt}{d\tau} = \frac{1}{\sqrt{1 - v^2}} \ \ .
\end{equation}
The scattering takes place in the plane $z = 0$ so that the spatial distance between
the incoming particle and the target is
\begin{equation}
R\left( \tau \right) = \left\vert \mathbf{x}\right\vert
= \sqrt{x^{2} +y^{2} }=\sqrt{\left( v\dot{t}_{in}\tau \right) ^{2}+s_{y}^{2}} \ \ .
\end{equation}
It is convenient take the correlation length $\lambda$ to be small so that the
support of the fields is narrowly centered around the retarded time
$\tau_1$.  Taking $\lambda \approx R(\tau_1)$ allows us to approximate $\varphi(\tau - \tau_1)
\approx \delta(\tau - \tau_1)$, so that $\tau_1$ is determined from the causality conditions
for the initial trajectories,
\begin{equation}
\left[ X_{in}\left( \tau _1\right) -X_{Z}\left( \tau _1\right)\rule[-2pt]{0pt}{10pt}\right] ^{2}=0
\qquad \qquad X^0_{in} \left(\tau_1 \right) - X^0_{Z}\left( \tau _1\right) > 0
\end{equation}
These equations have the solution
\begin{equation}
\tau _1=\frac{1}{v\dot{t}_{in}\left( 1-\eta_{v}^{2}\right) }\left( \eta
_{v}s_{t}+\sqrt{s_{t}^{2}-s_{y}^{2}\left( 1-\eta_{v}^{2}\right) }\right) 
\xrightarrow[\,\,\,   v \, \ll \, 1 \,\,\,  ]{}\frac{\sqrt{s_{t}^{2}-s_{y}^{2}}}{v}
\end{equation}
where it is convenient to introduce the smooth parameter
\begin{equation}
\eta_{v} = \frac{1}{v}\left( 1-\frac{1}{\dot{t}_{in} }\right) 
\longrightarrow \left\{
\begin{array}{ll}
0, & v = 0 \\
1, & v = 1 
\end{array}
\right. \ \ .
\end{equation}
Notice that the 0-component $s_t$ of the impact parameter must be positive in order
for the interaction to take place.
The location of the incoming particle at the time of interaction is found to be
\begin{equation}
\mathbf{x}\left( \tau _1\right)
=R \, \mathbf{\hat{R}} \qquad \qquad 
t\left( \tau _1\right)
= \dot{t}_{in} \tau_1 + s_{t}
\end{equation}
where
\begin{equation}
R=\frac{1}{1-\eta_{v}^{2}}\left( \eta_{v}\sqrt{s_{t}^{2}-s_{y}^{2}\left(
1-\eta_{v}^{2}\right) }+s_{t}\right) \xrightarrow[\,\,\,   v \, \ll \, 1 \,\,\,  ]{}s_{t}
\end{equation}

\begin{equation}
\mathbf{\hat{R}}=\frac{\left( \eta_{v}s_{t}+\sqrt{s_{t}^{2}-s_{y}^{2}\left( 1-\eta
_{v}^{2}\right) },\left( 1-\eta_{v}^{2}\right) s_{y},0\right) }{s_{t}+\eta
_{v}\sqrt{s_{t}^{2}-s_{y}^{2}\left( 1-\eta_{v}^{2}\right) }}
\xrightarrow[\,\,\,   v \, \ll \, 1 \,\,\,  ]{}
\left( \sqrt{1-\frac{s_{y}^{2}}{s_{t}^{2}}},\frac{s_{y}}{s_{t}},0\right) \ \ .
\end{equation}

From equation (\ref{yukawa}) the potential induced by the target nucleus is
\begin{equation}
a^{0}\left( x,\tau \right) =a^{5}\left( x,\tau \right) =\frac{ Ze}{4\pi R }
\delta \left( \tau -\tau_1 \right)
\qquad a^i = 0
\label{a0}
\end{equation}
so that the nonzero field strengths can be written
\begin{equation}
e^i = \partial^0 a^i - \partial^i a^0 \qquad 
f^{5i} = \partial^5 a^i - \partial^i a^5 \qquad 
f^{50} = \partial^5 a^0 - \partial^0 a^5
%
\end{equation}
\begin{equation}
\mathbf{e}=-\nabla a^{0} \qquad 
\mathbf{f^5 }=-\nabla a^{5}= \mathbf{e} 
\qquad f^{50} =-\left( 1 + \dfrac{1}{\dot{t}_{in}} \right) \partial _{\tau }a^{0} \ \ .
\end{equation}
Using these expressions in (\ref{lorentz}) provides the Lorentz force on the incoming
particle in the form
\begin{eqnarray}
\ddot{t} \eq -\frac{e_{0}}{M}\mathbf{e}\cdot \mathbf{\dot{x}}-\frac{e_{0}%
}{M}f^{50}
=\frac{\lambda e}{M} \left( \mathbf{\dot{x}} \cdot \nabla  +
\left( 1 + \dfrac{1}{\dot{t}_{in}} \right) \partial _{\tau } \right)
a^{0}\left( x,\tau \right) \strt{16}
\\
\mathbf{\ddot{x}} \eq -\frac{e_{0}}{M}
\mathbf{e}\dot{t}+\frac{e_{0}}{M}\mathbf{f^5 }
=\frac{\lambda e}{M} \big( \dot{t}+1 \big)\nabla a^{0}\left( x,\tau \right) \ \ .
\end{eqnarray}
The delta function in (\ref{a0}) enables immediate integration of the force equations
as
\begin{eqnarray}
  \dot{t}_f -  \dot{t}_{in} 
&=& \frac{\lambda}{M}  
\dint_{\tau _1-\lambda / 2 }^{\tau _1+\lambda / 2
}d\tau \; \left( \mathbf{\dot{x}} \cdot \nabla  +
\left( 1 + \dfrac{1}{\dot{t}_{in}} \right) \partial _{\tau } \right)
\frac{Z e^2}{4\pi R} \delta \left( \tau - \tau_1 \right) \notag\\
\strt{8}
&=&\frac{\lambda}{M} \;
\mathbf{\dot{x}}\left( \tau _1\right) \cdot \nabla \,
\frac{Z e^2}{4\pi R}
\strt{8}  \notag\\
&=& - \frac{\lambda}{M} \, \frac{Z e^2}{4\pi R^2} \;
\mathbf{\dot{x}}\left( \tau _1\right) \cdot
\mathbf{\hat{R}} \strt{18} \label{t-change} \\
\mathbf{\dot{x}}_f -\mathbf{\dot{x}}_{in}
&=&\frac{\lambda}{M}
\dint_{\tau _1-\lambda / 2 }^{\tau
_1+\lambda / 2 }d\tau \; \big(
\dot{t}+1 \big)\nabla \frac{Z e^2}{4\pi R}
\delta \left( \tau - \tau_1 \right)
\strt{18} \notag
\\
&=&- \frac{\lambda}{M} \, \frac{Z e^2}{4\pi R^2} 
\big(
\dot{t}\left( \tau _1\right)+1 \big) \, \mathbf{\hat{R} \label{x-change} }
\end{eqnarray}
where the velocities are evaluated at the interaction point as
\begin{equation}
\left(  \dot{t} , \mathbf{\dot{x}} \strt{4} \right)\left( \tau _1\right)
= \frac12 \left[
\left(   \dot{t} , \mathbf{\dot{x}} \right)_f + \left(   \dot{t} , \mathbf{\dot{x}}
\right)_{in}
\right] \ \ .
\end{equation}
We introduce the dimensionless parameter
\begin{equation}
g_e =\frac{\lambda}{M} \, \frac{Z e^2}{4\pi R^2}
=\frac{\lambda }{R} \times \frac{Ze^{2}}{4\pi R}\frac{1}{M}
=\frac{{\text{correlation length}}}{{\text{impact parameter}}}
\times \frac{{\text{interaction energy}}}{{\text{mass energy}}}
\label{ge}
\end{equation}%
which appears in (\ref{t-change}) and (\ref{x-change}) as the factor controlling the
strength of the interaction.  Writing
\begin{equation}
\alpha_x = \frac{1}{2}g_e \hat{R}_{x} \qquad \qquad \alpha_y = \frac{1}{2}g_e \hat{R}_{y}
\end{equation}
we can expand the Lorentz force as components in the form
\begin{equation}
\left[ 
\begin{array}{ccc}
1 & \alpha_x & \alpha_y \\ 
\alpha_x & 1 & 0 \\ 
\alpha_y & 0 & 1%
\end{array}%
\right] \left[ 
\begin{array}{c}
\dot{t}_f  \\ 
\dot{x}_f  \\ 
\dot{y}_f 
\end{array}%
\right] =\left[ 
\begin{array}{ccc}
1 & -\alpha_x & 0 \\ 
-\alpha_x & 1 & 0 \\ 
-\alpha_y & 0 & 0%
\end{array}%
\right] \left[ 
\begin{array}{c}
\dot{t}_{in}  \\ 
v\dot{t}_{in} \\ 
0%
\end{array}%
\right] 
-2 \, \left[ 
\begin{array}{c}
0 \\ 
\alpha_x \\ 
\alpha_y%
\end{array}%
\right] 
\end{equation}
and solve for the final velocity
\begin{equation}
\left[ \hspace{-2pt}
\begin{array}{c}
\dot{t}_f \strt{8} \\ 
\dot{x}_f \strt{8} \\ 
\dot{y}_f 
\end{array}%
\hspace{-2pt}
\right] =\frac{1}{1-\frac{1}{4}g_e ^{2}}\left\{ \left[ \hspace{-2pt}
\begin{array}{c}
\dot{t}_{in} \strt{8} \\ 
v\dot{t}_{in} \strt{8} \\ 
0%
\end{array}%
\hspace{-2pt}
\right] -g_e \left[ \hspace{-4pt}
\begin{array}{c}
\dot{t}_{in}   v \hat{R}_{x}\strt{6}\\ 
\left( \dot{t}_{in} + 1 \right) \hat{R}_{x} \strt{8}\\ 
\left( \dot{t}_{in} + 1 \right) \hat{R}_{y}
\end{array}%
\hspace{-4pt}
\right] +\frac{1}{4}g_e ^{2}\left[ \hspace{-4pt}
\begin{array}{c}
\dot{t}_{in} +2  \strt{8}\\ 
\left( \hat{R}_{x}^{2}-\hat{R}_{y}^{2}\right) v\dot{t}_{in}  \strt{8}\\ 
2\hat{R}_{x}\hat{R}_{y}v\dot{t}_{in} 
\end{array}%
\hspace{-4pt}
\right] \right\} \ \ .
\label{gen-case}
\end{equation}
Before considering pair annhilation, we examine the low velocity and low
interaction energy limit of this result.  Taking
\begin{equation}
\left\vert \mathbf{\dot{x}}\right\vert = v \ll 1 \qquad\dot{t}_{in} \rightarrow 1 \qquad
\eta_{v} \rightarrow 0 \qquad g_e \ll 1 
\label{low_E}
\end{equation}
the initial velocity reduces to
\begin{equation}
\dot{X}_{in}\left( \tau \right) 
\rightarrow \left( 1,v,0,0\right) \ \ ,
\end{equation}
the final velocity becomes
\begin{equation}
\dot{t}_f \approx \dot{t}_{in} \qquad \qquad 
\mathbf{\dot{x}}_f  \approx \mathbf{\dot{x}}
- 2 g_e \mathbf{\hat{R}} 
\qquad \qquad
\mathbf{\hat{R}} = 
\left( \sqrt{1-\frac{s_{y}^{2}}{s_{t}^{2}}},\frac{s_{y}}{s_{t}},0\right)
\qquad \qquad
R = s_t
\end{equation}
%
and the scattering angle can be found as
%
\begin{equation}
\cos \theta =\frac{\mathbf{\dot{x}}_f
\cdot \mathbf{\dot{x}}}{\left\vert \mathbf{\dot{x}}_f \right\vert
\left\vert \mathbf{\dot{x}}\right\vert }
=\frac{\mathbf{\dot{x}}^2 -2 g_e  \mathbf{\hat{R}}
\cdot \mathbf{\dot{x}}}
{\left\vert \mathbf{\dot{x}}_f \right\vert
\left\vert \mathbf{\dot{x}}\right\vert }
=\frac{v -2 g_e  \hat{R}_x}
{\left\vert \mathbf{\dot{x}}_f \right\vert } \ \ .
\end{equation}
If we impose the nonrelativistic condition for conservation of energy, we obtain a
new constraint in the form
\begin{equation}
\mathbf{\dot{x}}^{2} = v^{2} = \mathbf{\dot{x}}^{2}_f =
\left[ \mathbf{\dot{x}}
- 2g_e  \mathbf{\hat{R}} \right]^2
\quad \Rightarrow \quad
v \hat{R}_x  =  g_e 
\end{equation}
in which case
\begin{equation}
\cos \theta =\frac{1}
{\left\vert \mathbf{\dot{x}}_f \right\vert }
\left[ v -g_e  \hat{R}_x \right]
=1 - 2 \hat{R}_x^2
\end{equation}
Now, using (\ref{ge}) we find
\begin{equation}
\cot \frac{\theta }{2} =\sqrt{\frac{1+\cos \theta }{1-\cos \theta }}= \frac{\hat{R}_y}{\hat{R}_x}
= \frac{s_{y}}{s_{t}} \;
 \frac{v}{g_e} 
=  \frac{s_t}{\lambda v} \times \frac{4\pi Mv^2 s_{y}}{Ze^{2}}
\end{equation}
which recovers the Rutherford scattering formula if
\begin{equation}
\frac{s_t}{\lambda v} =  1 \ \ .
\label{ruth_cond}
\end{equation}
But from (\ref{low_E}) we have $s_t = R\left(\tau_1\right)$ which we assumed to be
comparable to $\lambda$.  Since $\lambda v  \ll \lambda$ in this low velocity case,
(\ref{ruth_cond}) cannot be maintained.  This result is unsurprising because the
short-range potential cannot provide an adequate model of nonrelativistic
Rutherford scattering. 

Removing these restrictions and returning to the relativistic case, the condition for
pair annihilation is that particle-1 scatters to negative energy, that is
$\dot{t}_f < 0$ for some value of $g_e$. From (\ref{t-change})
\begin{equation}
\dot{t}_f =\frac{\dot{t}_{in} \left( 1-g_e v
\hat{R}_{x}\right) +\frac{1}{4}g_e ^{2}\left( \dot{t}_{in} +2\right) }{1-\frac{1}{4}g_e ^{2}}
%
\end{equation}
and we see that for small values of $g_e$,
\begin{equation}
\dot{t}_f \longrightarrow \dot{t}_{in} \ge 1 \ \ .
\end{equation}
Since $v <1$ and $R_{x} < 1$, the discriminant of the numerator satisfies
\begin{equation}
\left( v R_{x}\right) ^{2} - \left[ 1+\frac{2}{\dot{t}_{in} } \right] <0
\end{equation}
so that the numerator is positive definite.
The denominator becomes negative when 
\begin{equation}
1-\frac{1}{4}g_e ^{2}<0\quad \Rightarrow \quad g_e
=\frac{{\text{correlation length}}}{{\text{impact parameter}}}
\times \frac{{\text{interaction energy}}}{{\text{mass energy}}} >2
\end{equation}
and since we take the correlation length $\lambda$ approximately equal to the 
impact parameter $R$, the requirement for pair annihilation is
\begin{equation}
\frac{Ze^{2}}{4\pi R} > 2 \, M
\end{equation}
meaning that the interaction energy is greater than the mass energy of the
annihilated particles.  As $g_e$ approaches 2 from below $\dot{t}_f$ becomes
very large.  After $g_e$ passes this critical value, $\dot{t}_f$ decreases from large
negative values, taking the limiting value
\begin{equation}
\dot{t}_f \underset{g_e \rightarrow
\infty}{\xrightarrow{\hspace*{1.5cm}}
} -\left( \dot{t}_{in} +2\right) \qquad \Longrightarrow \qquad 
E_f = -(E_{in} + 2M)
\end{equation}
so that the outgoing trajectory is timelike for all values of $g_e$.

\section{Bremsstrahlung}

Having found the condition for pair annihilation at time $\tau_1$ we now apply the
general expression (\ref{gen-case}) to describe the scattering at time $\tau_2$.
Particle-2 approaches a second nucleus along some trajectory $x^\mu_{in} (\tau)$ and
emerges from the interaction along trajectory $x^\mu_f(\tau)$ with positive energy.
To find the radiation field emitted by the
scattering and acceleration of particle-2, 
at a point $y^\mu$ along a line of observation
\begin{equation}
z = y - x(\tau_2) \ \ ,
\end{equation}
we write the initial and final 4-velocities as  
\begin{equation}
u_{in}= \dot{x}_{in} \qquad \qquad  u_f = \dot{x}_f 
\end{equation}
so that
\begin{eqnarray}
\Delta u =&&\hspace{-14pt}  u_{f}-u_{in} \\
u\left( \tau \right)  =&&\hspace{-14pt}  u_{in}+\Delta u~\theta \left( \tau -\tau _2\right) \\
\dot{u} \left( \tau \right) =&&\hspace{-14pt} \Delta u~\delta \left( \tau -\tau _2\right) \\
u \left( \tau_2 \right) =&&\hspace{-14pt} \bar{u} = \frac12\left[ u_{f}+u_{in}\right]
\ \ .
\end{eqnarray}
From (\ref{rad-1}) and (\ref{rad-2}) we rewrite the radiation fields produced by an
arbitrary trajectory in the form
\begin{eqnarray}
f_{rad}^{\mu \nu } \eq -e\varphi (\tau -\tau_2){\cal F}^{\mu\nu}
\left(z,u,\dot{u}\right)
- e \varphi^\prime (\tau -\tau_2){\cal G}^{\mu\nu}\strt{12}
\\
f_{rad}^{5\mu }  \eq e\varphi (\tau -\tau_2){\cal F}^{5\mu}
\left(z,u,\dot{u}\right)
- e \varphi^\prime (\tau -\tau_2){\cal G}^{5\mu}
\end{eqnarray}
where 
\begin{equation}
{\cal F}^{\mu \nu } = \left[\frac{\left( z\wedge \dot{u}\right) \left( u\cdot z\right)
-\left( z\wedge u\right) \left( \dot{u}\cdot z\right) }{{4\pi }%
\left( u\cdot z\right) ^{3}}\right]^{\mu\nu}
\qquad 
{\cal G}^{\mu \nu } = {\frac{1}{{4\pi }}}\left[ \frac{z\wedge u}{\left( u\cdot z\right) ^{2}}\right]^{\mu \nu } \strt{24}
\end{equation}
\begin{equation}
{\cal F}^{5\mu }  = \left[\frac{\left( \dot{u}\cdot z\right) z}{{4\pi }\left( u\cdot
z\right) ^{3}}\right]^{\mu }
\qquad 
{\cal G}^{5\mu } = \left[\frac{z-u\left( u\cdot z\right) }{{%
4\pi }\left( u\cdot z\right) ^{2}}\right]^{\mu } \vspace{8pt} \ \ .
\end{equation}

As pictured in Figure 4, the radiation emitted by the scattering of particle-2 is
absorbed by the negative energy particle-1 arriving at $y^\mu$.  Using
the Lorentz force equations (\ref{lorentz}) we calculate the change in velocity to
particle-1 caused by the incoming radiation.
Since we take $\lambda$ to be small, we may approximate the smoothing kernel as
\begin{equation}
\varphi \left( \tau - \tau_2 \right) =\dfrac{1}{2\lambda }\left[ \theta \left( \tau
-\strt{04.5}\left(  \tau_2 -\lambda \right) \right)
-\theta \left( \tau -\strt{04.5}\left(  \tau_2 +\lambda 
\right) \right) \right] \ \ .
\end{equation}
The $\tau$ integrations over $\varphi^\prime (\tau -\tau_2)$ vanish, leaving
the change in velocity $\dot{y}^{\mu }(\tau_2)$ of particle-1 in the form
\begin{eqnarray}
\Delta\dot{y}^{\mu } \eq \frac{\lambda e}{M} \int_{-\infty }^{\infty }d\tau ~
\left[f^{\mu\nu}_{rad} \dot{y}_\nu 
+ f^{\mu 5}_{rad} \dot{y}_5\right] 
= -\frac{e^2}{2M} \int_{\tau _{2}-\lambda }^{\tau
_{2}+\lambda }d\tau ~
\left[{\cal F}^{\mu\nu} \dot{y}_\nu 
+ {\cal F}^{\mu 5} \dot{y}_5\strt{08}\right] \notag
\\
\eq -\frac{e^2}{2M} 
\left[{\cal F}^{\mu\nu} \left(z,\bar{u},\Delta u\right)\dot{y}_\nu 
+ {\cal F}^{5 \mu} \left(z,\bar{u},\Delta u\right)\strt{08}\right] 
\label{lor-rad}
\end{eqnarray}
expressed in terms of the velocity change $\Delta u$ and average velocity $\bar{u}$,
which are found from (\ref{gen-case}) to be
\begin{equation}
\Delta u = -\frac{g_e }{1-\frac{1}{4}g_e ^{2}}\left[ 
\begin{array}{c}
v\dot{t}_{in}\hat{R}_{x}  \strt{8} \\ 
\left( \dot{t}_{in} +1 \right) \hat{R}_{x}  \strt{8} \\ 
\left( \dot{t}_{in} +1 \right) \hat{R}_{y}%
\end{array}
\right] +\frac{\frac{1}{2}g_e ^{2}}{1-\frac{1}{4}g_e ^{2}}\left[ 
\begin{array}{c}
\left( \dot{t}_{in} +1 \right)  \strt{8}  \\ 
v\dot{t}_{in}\hat{R}_{x}^{2} \strt{8}  \\ 
v\dot{t}_{in}\hat{R}_{x}\hat{R}_{y}
\end{array}
\right] 
\end{equation}\vspace{-4pt}\\
\begin{equation}
\bar{u}=
\frac{1}{1-\frac{1}{4}g_e ^{2}}\left[ 
\begin{array}{c}
\dot{t}_{in}  \\ 
v\dot{t}_{in}  \\ 
0%
\end{array}%
\right] -\frac{\frac{1}{2}g_e }{1-\frac{1}{4}g_e ^{2}}\left[ 
\begin{array}{c}
v\dot{t}_{in}\hat{R}_{x} \strt{8}  \\ 
\left( \dot{t}_{in} + 1 \right) \hat{R}_{x} \strt{8}  \\ 
\left( \dot{t}_{in} + 1 \right) \hat{R}_{y}%
\end{array}%
\right] +\frac{\frac{1}{4}g_e ^{2}}{1-\frac{1}{4}g_e ^{2}}
\left[ 
\begin{array}{c}
1 \strt{8}  \\ 
-v\dot{t}_{in}\hat{R}_{y}^{2} \strt{8}  \\ 
v\dot{t}_{in}\hat{R}_{x}\hat{R}_{y}%
\end{array}%
\right] \ \ .
\end{equation}

Since the support of $\varphi(\tau - \tau_2)$ is narrowly centered on $\tau_2$, the
line of observation $z^\mu$ must be a lightlike vector, which we write as
\begin{equation}
z^\mu  = y^\mu - x^\mu \left(\tau_2\right)= \rho \hat{n}^\mu
\quad \quad 
\hat{n} = \left(1,\mathbf{\hat{n}}\right) \;\;,\;\; \mathbf{\hat{n}}^2 = 1 \ \ .
%
\end{equation}
From (\ref{lor-rad}) the Lorentz force acting on particle-1 at $\tau_2$ can be
written
\begin{eqnarray}
\dot{y}^{\mu }_f +\frac{e^{2}}{4M}%
&&\hspace{-18pt}\frac{\left( z^{\mu }\Delta u_{\nu }-z_{\nu }\Delta u^{\mu }\right) \left( 
\bar{u}\cdot z\right) -\left( z^{\mu }\bar{u}_{\nu }-z_{\nu }\bar{u}^{\mu
}\right) \left( \Delta u\cdot z\right) }{{4\pi }\left( \bar{u}\cdot z\right)
^{3}}\dot{y}^{\nu }_f  \strt{12} \notag \\
&&\hspace{-18pt}=\dot{y}^{\mu }_i -\frac{1}{4}\frac{e^{2}}{M
}\frac{\left( z^{\mu }\Delta u_{\nu }-z_{\nu }\Delta u^{\mu }\right) \left( 
\bar{u}\cdot z\right) -\left( z^{\mu }\bar{u}_{\nu }-z_{\nu }\bar{u}^{\mu
}\right) \left( \Delta u\cdot z\right) }{{4\pi }\left( \bar{u}\cdot z\right)
^{3}}\dot{y}^{\nu }_i \strt{12} \notag \\
&&\quad +\frac{e^{2}
}{2M}\frac{\left( \Delta u\cdot z\right) z^{\mu }}{{4\pi }\left( \bar{u}%
\cdot z\right) ^{3}}
\end{eqnarray}
Making the simplifying choice $\mathbf{\hat{R}}\cdot \mathbf{\hat{n}} =0$, we find
\begin{equation}
\bar{u}\cdot z 
=-\rho \gamma \left[ 1-v \left( \hat{n}_{x}+\frac{1}{2} g_e
\hat{R}_{x}\right) \right]
\qquad 
\Delta u\cdot z=g_e \gamma v\rho \hat{R}_{x} 
\end{equation}
so that taking $v \ll 1$ and neglecting $g_e^2$, the Lorentz force equations become
\begin{equation}
\dot{y}_{f}^{0}-g_e g_R
\mathbf{\hat{R}}\cdot \mathbf{y}_{f}=\dot{y}
_{in}^{0}+g_e g_R
\mathbf{\hat{R}}\cdot \mathbf{\dot{y}}_{in}
\end{equation}
\begin{equation}
\mathbf{\dot{y}}_{f}+g_e g_R
\left[ \left( \mathbf{\hat{n}}\cdot \mathbf{\dot{y}}_{f}-\dot{y}^{0}_{f}\right)
\mathbf{\hat{R}}-\mathbf{\hat{n}}\left( \mathbf{\hat{R}}
\cdot \mathbf{\dot{y}}_{f}\right) \right]
=\mathbf{\dot{y}}_{in}-g_e g_R
\left[ \left( \mathbf{\hat{n}}\cdot \mathbf{\dot{y}}
_{in}-\dot{y}^{0}_{in}\right) \mathbf{\hat{R}}-\mathbf{\hat{n}}
\left( \mathbf{\hat{R}}\cdot \mathbf{\dot{y}}_{in}\right) 
\right] 
\end{equation}
where
\begin{equation}
g_R = \frac{1}{2M}\frac{e^{2}}{{4\pi } \rho } \ \ \strt{18} .
\label{g_R}
\end{equation}
We write the velocity of incoming negative energy particle-1 as
\begin{equation}
\dot{y}_{in}^{0} < -1
\qquad \qquad 
\mathbf{\dot{y}}_{in}\cdot \mathbf{\hat{n}} = 0 \;\; \Rightarrow \;\;
\mathbf{\dot{y}}_{in}=\left\vert \mathbf{\dot{y}}_{in}\right\vert \mathbf{\hat{R}}
\end{equation}
and write the Lorentz force in components, with $g = g_e g_R$, as
\vspace{6pt}
\begin{equation}
\left[ \hspace{-4pt}
\begin{array}{ccc}
1 & -g \hat{R}_{x} & -g \hat{R}_{y} \strt{8}\\ 
-g \hat{R}_{x} & 1-g \hat{n}_{x}\hat{R}_{x} & -g \hat{n}_{x}
\hat{R}_{y} \strt{8}\\ 
-g \hat{R}_{y} & -g \hat{n}_{y}\hat{R}_{x} & 1-g \hat{n}_{y}
\hat{R}_{y}
\end{array}
\hspace{-4pt}\right] \left[ \hspace{-2pt}
\begin{array}{c}
\dot{y}_f^{0} \strt{8}\\ 
\dot{y}_{xf} \strt{8}\\ 
\dot{y}_{yf}
\end{array}
\hspace{-2pt}\right] = \left[\hspace{-2pt} 
\begin{array}{ccc}
1 & 0 & 0 \strt{8}\\ 
g \hat{R}_{x} & 1 & 0 \strt{8}\\ 
g \hat{R}_{y} & 0 & 1
\end{array}
\hspace{-2pt}\right] \left[ \hspace{-2pt}
\begin{array}{c}
\dot{y}_i^{0} \strt{8}\\ 
\dot{y}_{xi} \strt{8}\\ 
\dot{y}_{yi}
\end{array}
\hspace{-2pt}\right] 
+g \left\vert \mathbf{\dot{y}}_i\right\vert \left[ \hspace{-2pt}
\begin{array}{c}
1 \strt{8}\\ 
\hat{n}_{x} \strt{8}\\ 
\hat{n}_{y}
\end{array}
\hspace{-2pt}\right] 
\vspace{6pt}
\end{equation}
so that the final velocity of particle-1 after absorbing the radiation is
\begin{eqnarray}
\left[ \hspace{-2pt}
\begin{array}{c}
\dot{y}_{f}^0 \strt{8}\\ 
\mathbf{\dot{y}}_{f}%
\end{array}%
\hspace{-2pt}\right]  =
\frac{1}{1-g ^{2}}\left[ \hspace{-2pt}
\begin{array}{c}
\dot{y}_{in}^{0} \strt{8}\\ 
\left\vert \mathbf{\dot{y}}_{in}\right\vert \mathbf{\hat{R}}%
\end{array}%
\hspace{-2pt}\right] &&\hspace{-18pt}+\frac{2g }{1-g ^{2}}\left[ 
\begin{array}{c}
\left\vert \mathbf{\dot{y}}_{in}\right\vert  \strt{8}\\ 
\dot{y}_{in}^{0}\mathbf{\hat{R}}+\left\vert \mathbf{\dot{y}}_{in}\right\vert \mathbf{\hat{n}}%
\end{array}%
\hspace{-2pt}\right] \rule[-20pt]{0pt}{4pt}
\notag \\
+&&\hspace{-18pt}\frac{g ^{2}}{1-g ^{2}}\left[ \hspace{-2pt}
\begin{array}{c}
\dot{y}_{in}^{0}+\left\vert \mathbf{\dot{y}}_{in}\right\vert  \strt{8}\\ 
2\dot{y}_{in}^{0}\mathbf{\hat{n}}+2\left\vert \mathbf{\dot{y}}_{in}\right\vert \mathbf{\hat{R}}%
\end{array}%
\hspace{-2pt}\right] 
+\frac{g ^{3}}{1-g ^{2}}
\left[ \hspace{-2pt}
\begin{array}{c}
\left\vert \mathbf{\dot{y}}_{in}\right\vert  \strt{8}\\ 
\left\vert \mathbf{\dot{y}}_{in}\right\vert \mathbf{\hat{n}}%
\end{array}%
\hspace{-2pt}\right] \ \ .
\end{eqnarray}%
The 0-component is
\begin{equation}
\dot{y}_{f}^{0}=\frac{1+g^{2}}{1-g^{2}}\dot{y}_{in}^{0}+g\frac{2+g+g^{2}}{1-g^{2}}%
\left\vert \mathbf{\dot{y}}_{in}\right\vert
%
\end{equation}
approximated at low velocity as
\begin{equation}
\dot{y}_{f}^{0}\approx \frac{1+g^{2}}{1-g^{2}} \ \dot{y}_{in}^{0} =  -\alpha \dot{y}_{in}^{0}
\end{equation}
where
\begin{equation}
g^{2} =\frac{ \alpha+1 }{ \alpha-1 } \quad \Rightarrow \quad \alpha = -\frac{1+g^{2}}{1-g^{2}}
\end{equation}
is written so that $\alpha > 1$ for a positive energy timelike particle. The exact
final velocity of the scattered particle is 
\begin{eqnarray}
\left[ \hspace{-2pt}%
\begin{array}{c}
\dot{y}_{f}^{0} \strt{8}\\ 
\mathbf{\dot{y}}_{f}%
\end{array}%
\hspace{-2pt}\right] =&&\hspace{-18pt}-\frac{\alpha -1}{2}\left[ \hspace{-2pt}%
\begin{array}{c}
\dot{y}_{in}^{0} \strt{8}\\ 
\left\vert \mathbf{\dot{y}}_{in}\right\vert \mathbf{\hat{R}}%
\end{array}%
\hspace{-2pt}\right] +\sqrt{\alpha ^{2}-1} \left[ 
\begin{array}{c}
\left\vert \mathbf{\dot{y}}_{in}\right\vert  \strt{8}\\ 
\dot{y}_{in}^{0}\mathbf{\hat{R}}+\left\vert \mathbf{\dot{y}}%
_{in}\right\vert \mathbf{\hat{n}}%
\end{array}%
\hspace{-2pt}\right]
\strt{18} \notag \\
&&\hspace{-18pt}\qquad \qquad -\frac{\alpha +1}{2}\left[ \hspace{-2pt}%
\begin{array}{c}
\dot{y}_{in}^{0}+\left\vert \mathbf{\dot{y}}_{in}\right\vert  \strt{8}\\ 
2\dot{y}_{in}^{0}\mathbf{\hat{n}}+2\left\vert \mathbf{\dot{y}}%
_{in}\right\vert \mathbf{\hat{R}}%
\end{array}%
\hspace{-2pt}\right] -\frac{\alpha +1}{2}\sqrt{\frac{\alpha +1}{\alpha -1}}%
\left[ \hspace{-2pt}%
\begin{array}{c}
\left\vert \mathbf{\dot{y}}_{in}\right\vert  \strt{8}\\ 
\left\vert \mathbf{\dot{y}}_{in}\right\vert \mathbf{\hat{n}}%
\end{array}%
\hspace{-2pt}\right]
\end{eqnarray}
with 0-component 
\begin{equation}
\dot{y}_{f}^{0}=
-\alpha \dot{y}_{in}^{0}-\frac{\alpha +1}{2}\left[ 1+\frac{3-\alpha }{ \sqrt{\alpha
^{2}-1}}\right] \left\vert \mathbf{\dot{y}}_{in}\right\vert \ \ .
\end{equation}
A pair creation event is observed at $\tau_2$ for $\alpha > 1$ which requires that
$g^2 = g_e^2 g_R^2 > 1$.  In this case, $g_e$ is the interaction strength for the
scattering of particle-2 to positive energy, so $g_e <2$.  From (\ref{g_R}) the
requirement for pair creation is then
\begin{equation}
\frac{e^{2}}{{4\pi } \rho } > \frac{2M}{g_e} \ \ .
\end{equation}

\section{Conclusions}

In this paper we have shown that a classical equivalent of the Bethe-Heitler
mechanism is permitted in Stueckelberg-Horwitz electrodynamics.  Although
Stueckelberg proposed his model with the goal of providing such a description, the
calculation has not been previously carried out in detail.  The process begins at
$\tau_1$ with a pair annihilation event produced by the scattering of an incoming
particle in the Coulomb field of a target nucleus.  The general solution for the
change in velocity leads to a requirement for this pair process 
\begin{equation}
g_e = \dfrac{\lambda }{R} \times \dfrac{Ze^{2}}{4\pi R}\dfrac{1}{M} > 2
\quad \Rightarrow \quad \dfrac{\lambda }{R} \times \dfrac{Ze^{2}}{4\pi R} >2M 
\end{equation}
which is reasonable on relativistic grounds and can be seen as consistent with the
QED requirement of interaction energy greater than the total mass energy of the particle
pair.  The next stage in the mechanism is the scattering to positive energy ($g_e<2$
for this interaction) of a
second particle in the Coulomb field of another nucleus.  The outgoing 4-velocity is
found from the general solution for Coulomb scattering, and this allows us to
calculate the radiation emitted by the accelerating particle.  Using the Lorentz
force on the first particle produced by absorption of bremsstrahlung we find the
requirement on pair creation to be
\begin{equation}
g_e g_R > 1 \quad  \Rightarrow \quad  \dfrac{e^{2}}{{4\pi } \rho } > \dfrac{2M}{g_e}
\end{equation}
which is similarly reasonable on relativistic grounds.

Further work is required to obtain a realistic description of the classical
Bethe-Heitler mechanism.  To describe the interaction for the long range Coulomb
force, it is necessary to solve the nonlinear differential equations that arise
from the smoothing function
\begin{equation}
\varphi (\tau) = \frac{1}{2 \lambda} e^{-\vert \tau \vert / \lambda}
\end{equation}
with a large enough $\lambda$ to accurately model Rutherford scattering.  It is
also necessary to examine the mass transfer in the pair processes and
check mass conservation among particles and fields.






%
%
%

%
%
%
\end{document}